\documentclass[journal,twoside]{IEEEtran}
\usepackage{epstopdf}
\usepackage{amsmath,amsfonts}
\usepackage{algorithmic}
\usepackage{algorithm}
\usepackage{array}
\usepackage[caption=false,font=normalsize,labelfont=sf,textfont=sf]{subfig}
\usepackage{textcomp}
\usepackage{stfloats}
\usepackage{url}
\usepackage{verbatim}
\usepackage{graphicx}
\usepackage{authblk}
\usepackage{multicol}
\usepackage{fancyhdr}
\usepackage{cite}

\providecommand{\keywords}[1]
{
  \small	
  \textbf{\textit{Keywords---}} #1
}

\fancypagestyle{mypagestyle}{%
  \fancyhf{}% Clear header/footer
  \fancyhead[OC]{Mespinosa \emph{et. al}}% Author on Odd page, Centred
  \fancyhead[EC]{Intruders cooperatively interact with a wall into granular matter}% Title on Even page, Centred
  \fancyfoot[C]{\thepage}%
}
\title{Intruders cooperatively interact with a wall \\ into granular matter}
\author[a]{M. Espinosa$^*$}
\author[a]{V. L. D\'{\i}az$^*$} 
\author[a]{A. Serrano-Muñoz}
\author[a,1]{E. Altshuler}

\affil[a]{Group of Complex Systems and Statistical Physics, Physics Faculty, University of Havana, 10400 Havana, Cuba}
\affil[1]{Corresponding author: ealtshuler{@}fisica.uh.cu}
\affil[*]{Equally contributed to the article}

\begin{document}

\twocolumn[
  \begin{@twocolumnfalse} 

\maketitle

%\authorcontributions{Author contributions: E. A. designed research; A. S., E. A. and V. L. D. constructed experimental apparatus. M. E., V.L.D. and A.S. performed research; M. E., and E.A. analyzed data, M.E. performed numerical simulations; E.A. and M. E. wrote the paper.}
%\authordeclaration{The authors declare no competing financial interests.}
%\equalauthors{}
%\correspondingauthor{\textsuperscript{1}To whom correspondence should be addressed. E-mail: ealtshuler{@}fisica.uh.cu}

% Keywords are not mandatory, but authors are strongly encouraged to provide them. If provided, please include two to five keywords, separated by the pipe symbol, e.g:

%\markboth {Mespinosa \MakeLowercase{\textit{et al.}}: Intruders cooperatively interact with a wall into granular matter}{Mespinosa \MakeLowercase{\textit{et al.}}: Intruders cooperatively interact with a wall into granular matter}
%\thispagestyle{firststyle}
%\ifthenelse{\boolean{shortarticle}}{\ifthenelse{\boolean{singlecolumn}}{\abscontentformatted} {\abscontent}}{}

\begin{abstract}

When a cylindrical object penetrates granular matter near a vertical boundary, it experiences two effects: its center of mass moves horizontally away from the wall, and it rotates around its symmetry axis. Here we show experimentally that, if two identical intruders instead of one are released side-by-side near the wall, both effects are also detected. However, unexpected phenomena appear due to a cooperative dynamics between the intruders. The net horizontal distance traveled by the common center of mass of the twin intruders is much larger than that traveled by one intruder released at the same initial distance from the wall, and the rotation is also larger. The experimental results are well described by the Discrete Element Method, which reveals a further unexpected phenomenon: when four intruders are released as a column near a wall, they penetrate like a chain that ``bends away" from the wall so its lower end is very strongly repelled away from it. 
\end{abstract}

\keywords{Granular matter, Boundary effects, Intruder penetration, Sedimentation \vspace{10px}}

\end{@twocolumnfalse}
]

\thispagestyle{empty}

Due to the discrete nature of grains and their dissipative interactions, granular matter shows unexpected behaviours often reminding either a solid or a liquid \cite{le1996ticking,eggers1999sand,altshuler2003sandpile,shinbrot2004granular,martinez2007uphill,altshuler2008revolving}. In particular, the penetration of solid intruders into granular materials is a complex process where our intuition based on the observation of fluids and solids separately is of little help.
A large body of work dealing with the penetration of low-velocity projectiles into granular beds has developed over the last two decades or so \cite{uehara2003low,walsh2003morphology,boudet2006dynamics,katsuragi2007unified,goldman2008scaling,nelson2008projectile,pacheco2010cooperative,pacheco2011infinite,torres2012impact,ruiz2013penetration, brzinski2013depth,altshuler2014settling,clark2014collisional,sanchez2014note,de2016lift,viera2017note,bester2017collisional}. Most of those studies have focused on the penetration far from the limits of the granular container: intruder-wall interactions have rarely been studied. In 2012, Nelson {\it et al.} analyzed the effect of a vertical wall in the endpoint of spheres released near it, observing horizontal repulsion by the wall \cite{nelson2008projectile}. In 2012, Katsuragi studied the quasi-statically penetration of a spherical intruder into a cylindrical granular column, and finds scaling laws relating the drag force and the wall pressures \cite{katsuragi2012nonlinear}. In a recent contribution, we have fully characterized the repulsion of cylindrical objects near walls, and have reported that they also rotate \cite{Diaz-Melian2020}.

Here, we go a step further in the study of intruder-wall interactions. We experimentally show that two intruders released side-by-side from the wall separate from it  following a cooperative dynamics where the repulsion between the wall and intruder farther from it is ``boosted" by the presence of the one closer to the boundary. This effect is so strong, that the net horizontal distance traveled by the common center of mass of the twin intruders is around six times larger than that traveled by one intruder released at the same initial distance from the wall. The experimental results are well described by Discrete Element Method simulations (DEM), which were used to study the behavior of intruder ``quartets" released from two initial geometries: an horizontal line where the intruder at the far left is touching the wall, and a vertical line were all the intruders are touching the vertical wall. While the horizontal quartet behaves like a natural extension of the twins, the vertical quartet bends away from the wall, causing that the initially lowest intruder experiences a very strong repulsion from the wall.

\section*{Experimental and DEM simulation details}
\begin{figure}[!b]
\includegraphics[width=230px]{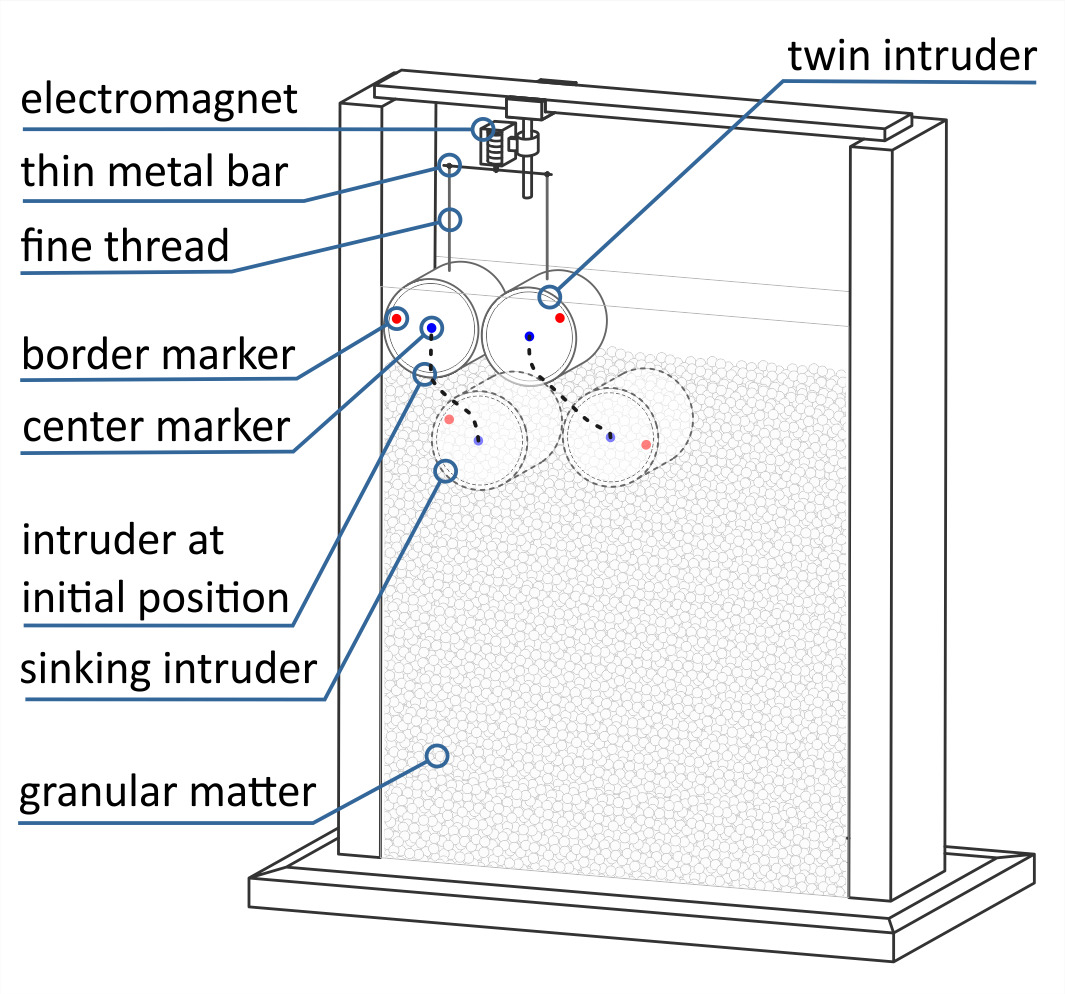}
%\vspace{3px}
%\includegraphics[width=230px]{Ac.eps}
%\centering
\caption{Experimental set up. Two identical cylinders were released into a Hele-Shaw cell as shown in the sketch. All trajectories resulting from the experiments were averaged over 10 measurements performed in the same conditions.}
\label{fig:Setup}
\end{figure}

Expanded polystirene spheres were deposited into a Hele-Shaw cell, as shown in Fig. \ref{fig:Setup} (the size distribution of the granular material and the dimensions of the cell were similar to those in \cite{Diaz-Melian2020}). Two cylindrical intruders of $3.75$~cm radius and $0.255$~kg mass were released from the surface of the granular bed by means of an electromagnetic device that minimized initial spurious vibrations and torques on the intruders. Initially, the left cylinder was gently touching the left vertical wall of the cell, and the right cylinder was gently touching the left cylinder, as illustrated in Fig. \ref{fig:Setup}. Videos of the penetration process were taken through one of the large faces of the cell using a digital camera. Two colored dots located at the center and near the border of one circular face of the cylinders served as reference points for image analysis \cite{reyes2021yupi}, so the motion of the individual intruders' centers of mass and rotation angles could be quantified.

We perform our numerical simulations using the discrete element model LAMMPS (Large-scale Atomic/Molecular Massively Parallel Simulator) \cite{plimpton2007lammps}, describing two identical cylindrical intruders sinking into a granular bed composed of spherical particles. The intruders are modeled using a $1$~mm-spacing simple cubic lattice, formed by spherical particles with $1$~mm diameter. The interaction between particles is ruled by a Hertzian contact model with a normal elastic constant $K_n = 3.6\times10^5$ and a tangential elastic constant $K_t = 5.5\times10^5$. We initiate each simulation by preparing the granular bed, pouring batches of particles with radius following a uniform random distribution between $1 - 3.25$~mm and a fixed density of $14\times10^{-3}$~g/cc. Each pour generates particles at random positions in a limited space of the container (cuboid with the same experimental dimensions) that moves in the $z$-axis as the container is filled, resulting in a total of $10^5$ particles. Both intruders are released simultaneously from the granular surface with zero initial velocity after the system relaxes. Fore more simulation details see the Supplemental Material.

\section*{Results and discussion}

\begin{figure}[!t]
\includegraphics[width=230px]{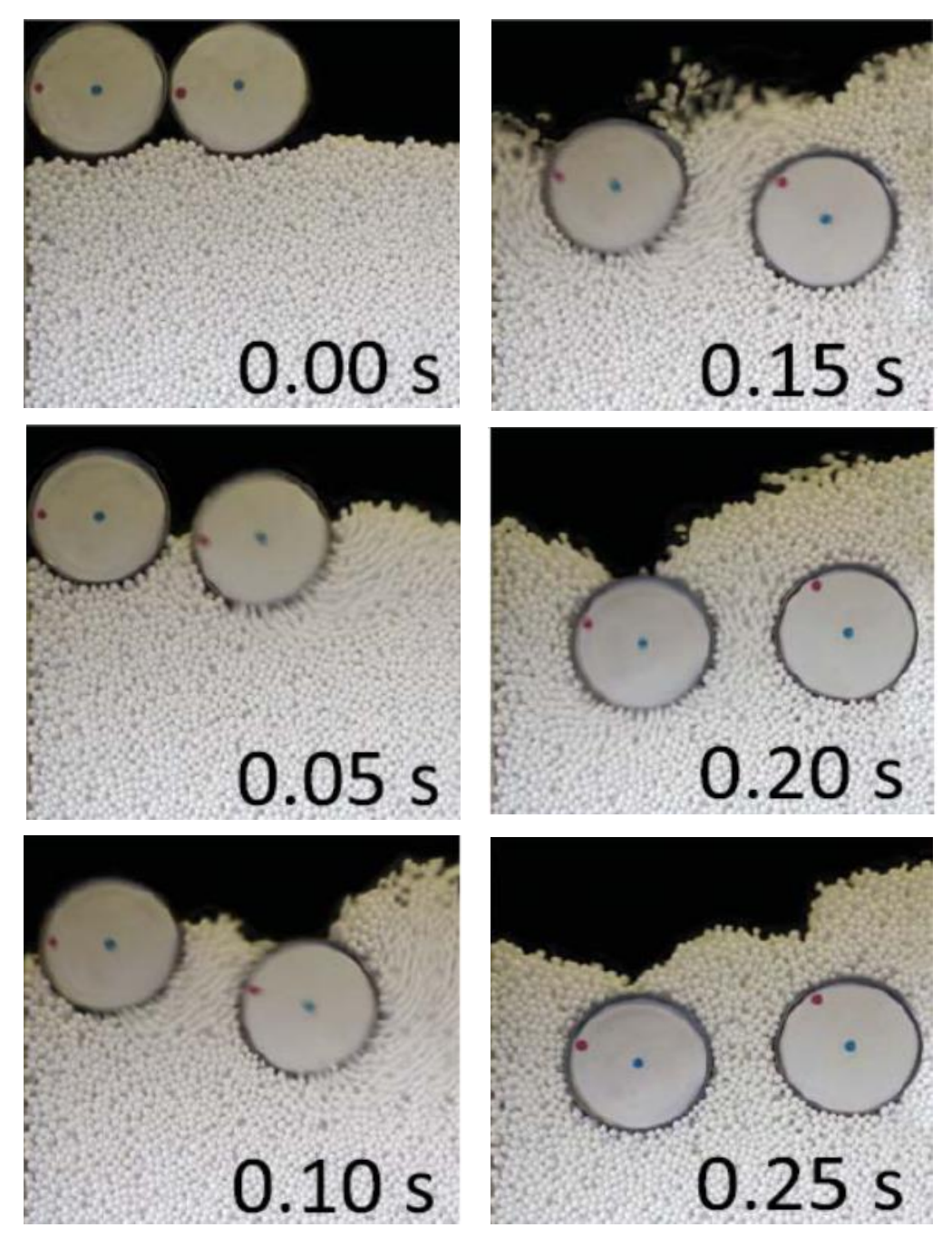}
%\vspace{3px}
%\includegraphics[width=230px]{Ac.eps}
%\centering
\caption{Cooperative dynamics of twin intruders moving away from the wall. Sequence of snapshots of the motion of two intruders released synchronously, in such a way that the one nearest the wall (NearTwin) touched it, and the one far from it (FarTwin) touched its NearTwin by its right edge (i.e., the left one was released at $x_0=0$\,cm and the right one at $x_0=7.5$\,cm). The left wall of the cell corresponds to the left boundary of each picture. Notice the clockwise rotation of both intruders.}
\label{fig:Snapshots}
\end{figure}

If two intruders are released side-by-side near a wall, a rich phenomenology is observed. Besides confirming that the intruders finish their motion at a nonzero horizontal distance from the wall (originally reported in \cite{nelson2008projectile}), we have been able to follow the whole repulsion process, which turns out to be cooperative in a non-trivial way. Fig. \ref{fig:Snapshots} shows a sequence of snapshots of the motion in one realization of the experiment that illustrates some of the main features of the motion (A sample video of one experiment is included as Supplemental Video 1). Up to approximately $0.1$~s after the release of the intruders, the one nearer to the wall (NearTwin) is almost frozen in space, while the one farther from the wall (FarTwin) penetrates the medium. This situation suggests that the right side of NearTwin is felt by FarTwin as an ``effective wall" located to the right of the real one. After approximately $0.1$~s, NearTwin starts to move, penetrating even deeper along the vertical direction than FarTwin. One can speculate that the latter has ``mobilized" the granular medium around NearTwin, facilitating its sinking. Finally, the red dots on each intruder clearly indicate that both rotate in the clockwise, or ``normal" direction \cite{Diaz-Melian2020}.

\begin{figure}[!b]
\includegraphics[width=260px]{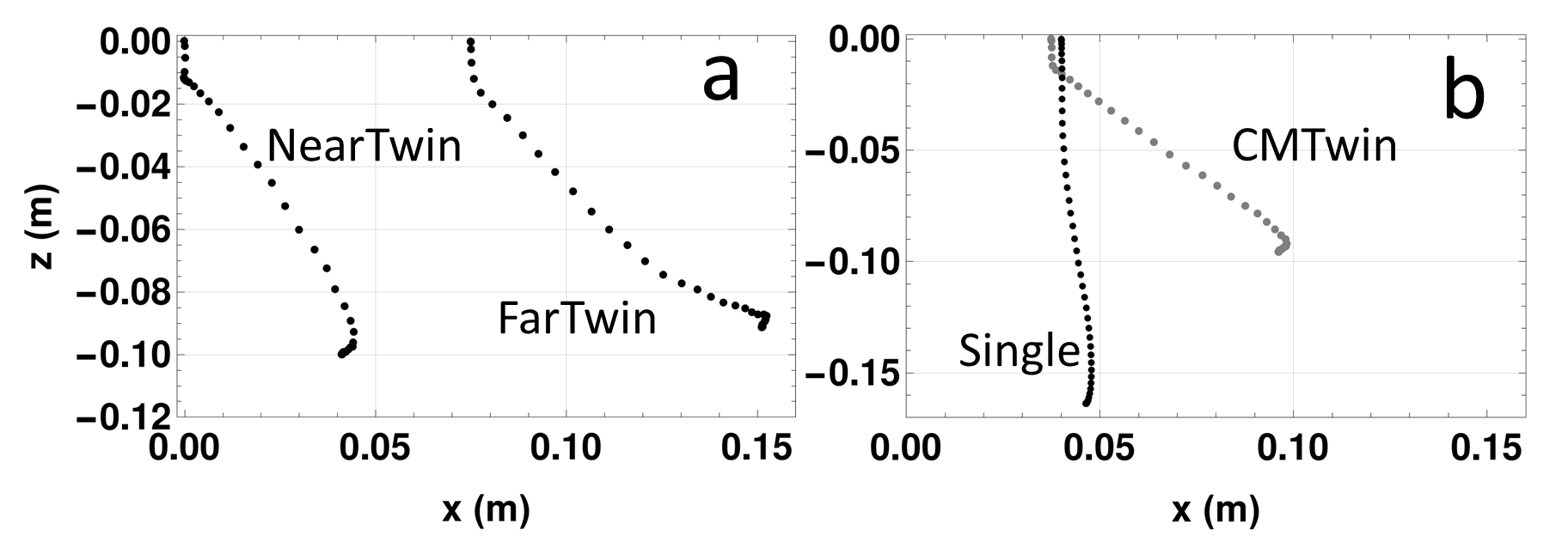}
%\vspace{3px}
%\includegraphics[width=230px]{Ac.eps}
%\centering
\caption{Trajectories. (a) Trajectories of the individual twin intruders after being released together. NeatTwin and FarTwin correspond to the intruders released near and far from the wall, respectively. (b) In gray, trajectory of the combined center of mass of the twin intruders (CMTwin). In black, trajectory of a single intruder (Single) released at an horizontal distance from the wall very near the analogous one for CMTwin. Trajectories averaged over 10 repetitions.}
\label{fig:Trajectories}
\end{figure}

Fig. \ref{fig:Trajectories} (a) indicates the trajectories of both intruders individually, resulting from the average over $10$ repetitions of the experiment. After an initial vertical plunge corresponding to an almost free fall, the granular material under the intruders is compacted: force chains are ``charged", preparing the conditions for both repulsion and rotation \cite{Diaz-Melian2020}. The gray dots in Fig.  \ref{fig:Trajectories} (b) represent the trajectory of the combined center of mass of the twin intruders (CMTwin), while the black dots correspond to experiments where a single intruder (Single) has been released from a lateral distance from the wall $x_0 = 4$~cm, ({\it i.e.}, approximately the initial position of CMTwin). The difference between both trajectories is dramatic: while the single intruder hardly ``feels" the presence of the wall (demonstrated by its almost vertical trajectory), the twin system experiences a large lateral repulsion --but also penetrates less in the vertical direction.

\begin{figure}
\includegraphics[width=255px]{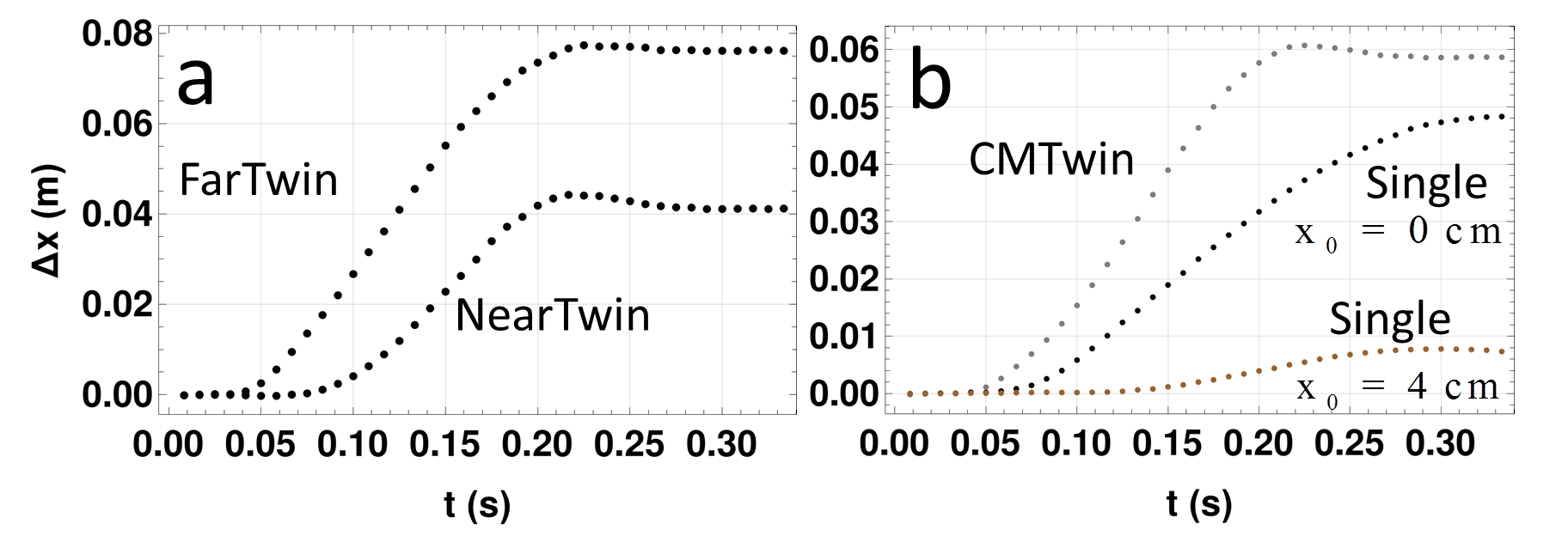}
%\vspace{3px}
%\includegraphics[width=230px]{Ac.eps}
%\centering
\caption{Net horizontal repulsion. (a) Time evolution of $\Delta x = x - x_0$ of the individual twin intruders after being released together. (b) In gray, time evolution of $\Delta x$ of the combined center of mass of the twin intruders. In black and brown, same graph for single intruders released at $x_0=0$ cm and $x_0=4$~cm from the wall, respectively. Trajectories averaged over $10$ repetitions.}
\label{fig:deltax}
\end{figure}

Fig. \ref{fig:deltax} illustrates how strong the lateral repulsion for the twin system is, as compared to the single intruder. In particular, Fig. \ref{fig:deltax} (b) reveals that the net horizontal motion of CMTwin is approximately 6 times larger than that for Single, which corresponds to a difference of nearly $5$~cm, equivalent to the final separation of a single intruder, when released at $x_0 = 0$~cm from the wall. This confirms the hypothesis that the right border of NearTwin plays the r\^{o}le of the lateral wall for FarTwin.

\begin{figure}[!b]
\includegraphics[width=255px]{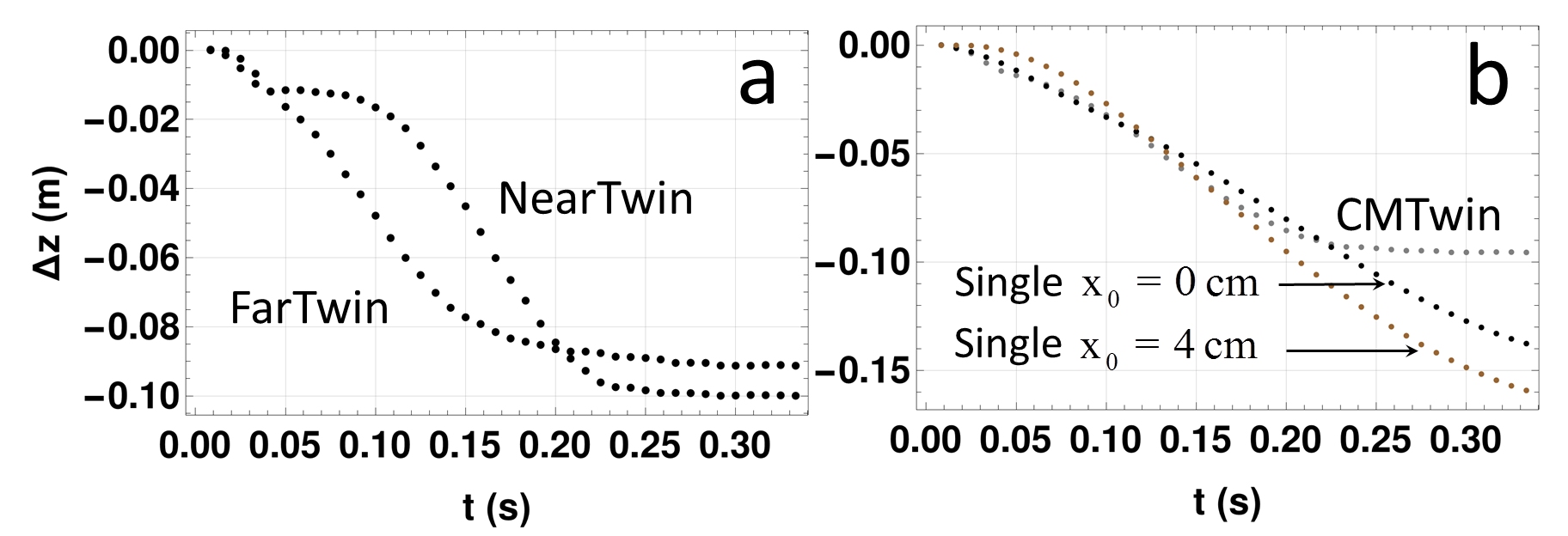}
%\vspace{3px}
%\includegraphics[width=230px]{Ac.eps}
%\centering
\caption{Vertical penetration. (a) Time evolution of $\Delta z = z - z_0$ of the individual twin intruders after being released together. (b) In gray, time evolution of $\Delta z$ of the combined center of mass of the twin intruders. In black and brown, same graph for single intruders released at a horizontal distance $x_0=0$ cm and $x_0=4$~cm from the wall, respectively. Trajectories averaged over $10$ repetitions.}
\label{fig:deltaz}
\end{figure}

Fig. \ref{fig:deltaz} (a) shows the vertical penetration of each of the twin intruders. The coincidence between both trajectories before approximately $0.04$~s corresponds to the vertical plunge seen in Fig. \ref{fig:Trajectories}, and the nearly constant position displayed by NearTwin between $0.04$~s and $0.10$~s corresponds to the situation suggested by Fig. \ref{fig:Snapshots}, where NearTwin is ``frozen", while FarTwin is moving. Fig. \ref{fig:deltaz} (b) shows a new unexpected result: before $0.2$~s, there is little difference in the sink process between CMTwin and Single, after which the twins basically stop moving downwards.

As suggested above, the cooperative motion of our twin intruders is based on their interaction mediated by the granular material. At the beginning of the penetration process, grains in the space between the cylinders move towards jamming, implying stronger force chains, which results in the large lateral repulsive motion of FarTwin while NearTwin is almost frozen. Subsequently they partially un-jam, facilitating the penetration of NearTwin afterwards. In order to gain insight into the nature of the interaction mechanisms, we have suppressed the grain-mediated interaction between the intruders in a DEM simulation where the twins are attached along a common line at their surfaces, resulting in a ``rigid" twin (RTwin). RTwin is released from a position indistinguishable from the one shown at the top left of Fig. \ref{fig:Snapshots}. Surprisingly, the center of mass of RTwin is horizontally repelled approximately the same amount as CMTwin, but RTwin penetrates less along the vertical axis. This study will be further explored in future work.

%quantifies the motion of the N and F twin intruders (as an average over 10 realizations of the experiment), and allow to compare them with their ``analogous" single intruders: those released at $x_0 = $ 0 and $x_0 = $ 7.5 cm, respectively. Figure \ref{fig:CenterOfMass}(b) demonstrates that the trajectory of the N intruder is quite similar to the single intruder released at $x_0 =$ 0 (red circles). However, N moves away farther from the wall during most of the trajectory, until it sharply ``returns" to the wall at the very end of the motion. Both features can be explained based on its interaction with F. As F moves away from the wall, it ``relaxes" the granular medium behind it, facilitating the motion of N. Nevertheless, when F stops, N ``bounces back" against F and moves slightly towards the wall. The motion of F itself departs dramatically from the trajectory of its single analogue released at $x_0 =$ 7 cm (brown curve). While the latter almost do not feel any effects from the wall, F moves horizontally away from the wall because the interaction with it is ``transmitted" by N. These features are corroborated by Fig. \ref{fig:CenterOfMass}(c), where the net horizontal motion is shown as time goes by. Interestingly, while the cooperative effects in the twin intruders enhance the repulsion from the wall, the penetration depth along the vertical is smaller than in the case of single intruders.

%\begin{SCfigure*}[\sidecaptionrelwidth][!ht]
%\includegraphics[width=330px]{D.eps}
%\centering \caption{ }
%\label{fig:FigD}
%\end{SCfigure*}

\begin{figure}[!b]
\includegraphics[width=255px]{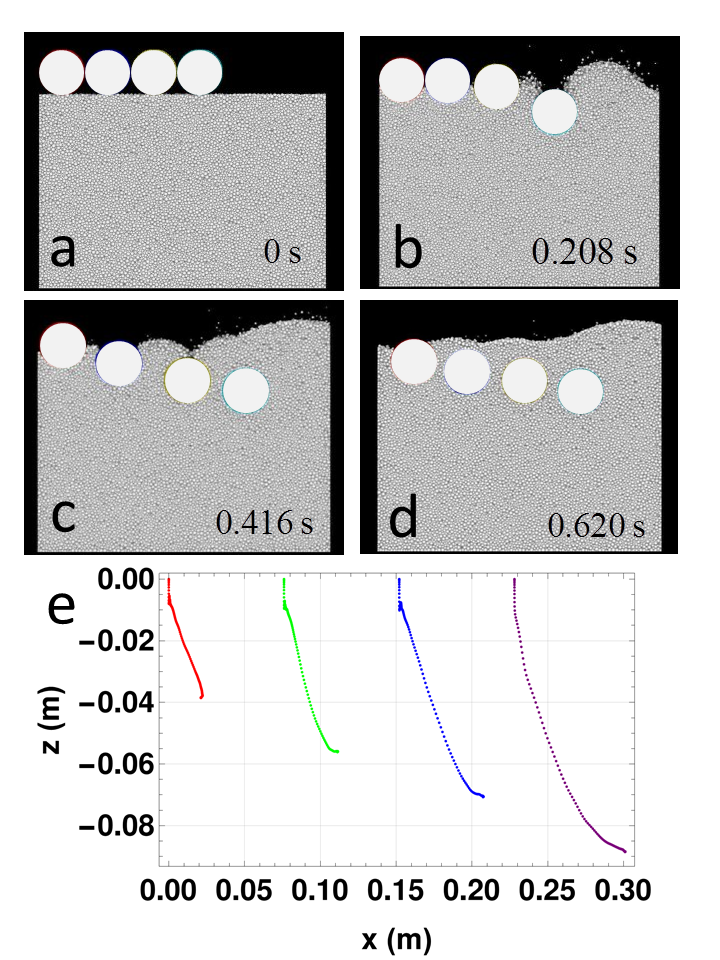}
%\vspace{3px}
%\includegraphics[width=230px]{Ac.eps}
%\centering
\caption{Horizontal quartet penetration. (a), (b), (c) and (d) show snapshots from DEM simulations taken at $0$~s, $0.208$~s, $0.416$~s and $0,620$~s, respectively. (e) Trajectories of the center of masses for all four intruders.}
\label{fig:FourHoriz}
\end{figure}

DEM simulations were run in order to mimic the behavior of the twin experiments described above. The simulations were able to reproduce very well the $\Delta x$ vs. $t$ and $\Delta z$ vs. $t$ behavior of the twins (See Supplemental Material), indicating that our simulations are an appropriate tool to investigate further experimental situations. Fig. \ref{fig:FourHoriz} show a series of snapshots taken from DEM simulations, in which four identical intruders (that we will call Quartet) are released in the horizontal arrangement shown in Fig. \ref{fig:FourHoriz} (a). The penetration dynamics looks like an extrapolation of the twin intruders: The one farther from the wall is rapidly repelled to the right, while the other ones are basically frozen; then the second, third and fourth intruders follow (An animation of the simulation is shown as Supplemental Video 2). The trajectories in Fig. \ref{fig:FourHoriz} (e) shows an initial vertical plunge of the four intruders, as well as an increasing horizontal repulsion and vertical penetration from the closest to the farthest intruder. If we launch a single intruder at a distance of $11.25$~cm from the wall (i.e., from the position of the Quartet's center of mass), it will not ``feel" the effect of the wall, with barely any lateral displacement, yet a much greater penetration depth ($\sim15$~cm).

In \cite{Diaz-Melian2020} we suggested the equivalence between two identical intruders released side-by-side far from the walls into a granular bed, and a couple of actual and image intruders, where the actual one is released from the wall --a picture reminding the method of images used in electrostatics and hydrodynamics \cite{blake1971note,altshuler2013flow}. Despite we cannot ensure that the ``image intruder" concept will facilitate the prediction of penetration trajectories near walls --even the theory of intruder-intruder interactions into granular matter is in its infancy-- it establishes evident links with the cooperative dynamics reported here. Indeed, Pacheco and Ruiz reported in 2010 the penetration dynamics of up to 5 disk-shaped intruders into a 2D Hele-Shaw cell filled with ultra light granular matter \cite{pacheco2010cooperative}. In particular, when they launched 5 intruders side-by-side (upper left corner of Fig. 3(a) in \cite{pacheco2010cooperative}), the center intruder stays behind at the beginning of the process while the rest penetrate and repel each other, forming an ``inverted V"-shaped flock. Then, the intruders at the edges slow down and the one at the center sinks faster, resulting in a ``V"-shaped flock. Finally, all 5 intruders land approximately at the same depth with a substantial increase in their lateral distances relative to their initial positions. If an imaginary vertical line is traced at the center of the picture, the disks at the right qualitatively behave as our twin intruders released close to a wall, considering that, in our case, the motion stops before developing the ``V" profile (we presume that the analogy would have been better if the authors had released 4 intruders instead of 5). Furthermore, Pacheco and Ruiz  observed that, as the number of intruders increases, the final penetration depth decreases \cite{pacheco2010cooperative} which is also evident in our case when we examine Fig. \ref{fig:deltaz}(b), and also in Fig. \ref{fig:FourHoriz}, corresponding to the horizontal quartet. These analogies strongly suggest that the image intruder concept can be expanded to multiple horizontal penetrators.

\begin{figure}[!t]
\includegraphics[width=255px]{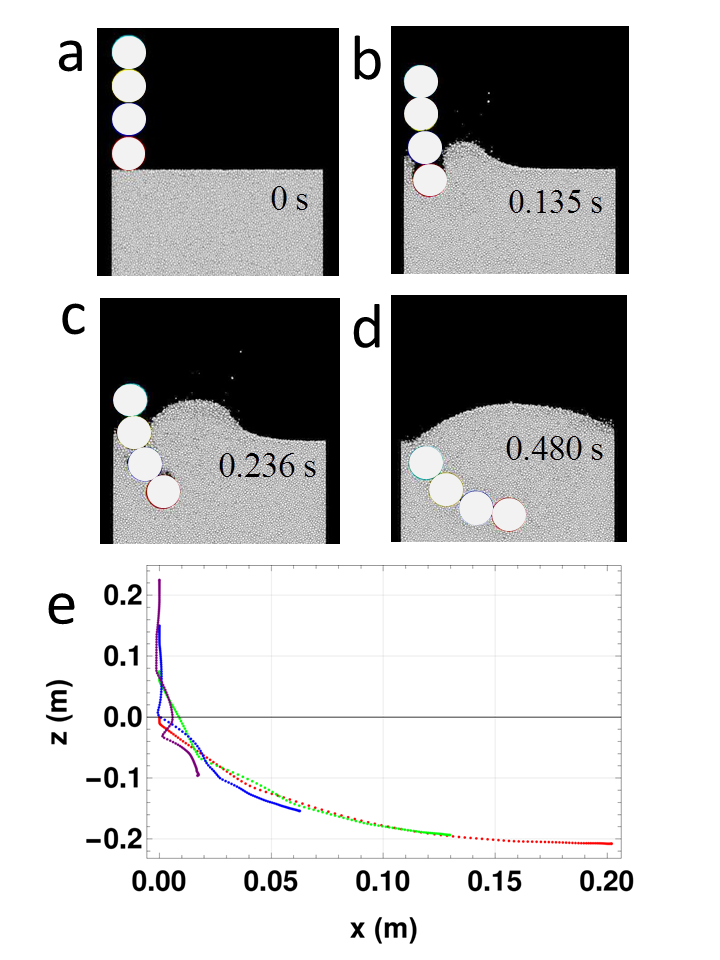}
%\vspace{3px}
%\includegraphics[width=230px]{Ac.eps}
%\centering
\caption{Vertical quartet penetration. (a), (b), (c) and (d) show snapshots from DEM simulations taken at $0$~s, $0.208$~s, $0.416$~s and $0,620$~s, respectively. (e) Trajectories of the center of masses for all four intruders.}
\label{fig:FourVert}
\end{figure}

However, the situation changes when a column-like quartet is released near the left wall, as depicted in Fig. \ref{fig:FourVert}. Differently from the case of various intruders released as a column far from the boundaries \cite{pacheco2010cooperative}, here the wall restrains the vertical motion of the intruders {\it before} they enter the granular material, avoiding the instability observed in Fig. 4 (b) of reference \cite{pacheco2010cooperative}. Then, we should not expect that the image intruder hypothesis works here. However, our simulation yields a quite unexpected result: differently from all the previous cases under study, the whole chain of intruders enters the granular bed in close contact, bending to the right due to repulsion from the wall, as illustrated in the snapshots shown in Fig. \ref{fig:FourVert} (a-d) (An animation of the DEM simulation is included as Supplemental Video 3). Then, the net motion along both the vertical and horizontal directions are much larger than those for the horizontal chains, which is easily appreciated by comparing Fig. \ref{fig:FourHoriz} (e) and \ref{fig:FourVert} (e). If we release a single intruder from a height of $11.25$~cm (corresponding to the initial position of the common center of mass of the vertical quartet) with $4$ times the intruder density, it manages to touch the bottom of the cell with a much smaller lateral displacement than the system of the $4$ vertical intruders. The same simulation, with the experimental density, reports a significantly smaller vertical penetration and horizontal separation from the wall. The large horizontal penetration can be explained qualitatively by observing the sequences of snapshots shown in Fig. \ref{fig:FourVert}. In (b), the force chains between the wall and the lower intruder force it to deviate to the right, so a large horizontal component of the weight of the upper three intruders pushes it further to the right. Qualitatively, the same situation repeats as the following two intruders penetrate the granular bed, resulting in a very large lateral repulsion.

\begin{figure}[!t]
\includegraphics[width=255px]{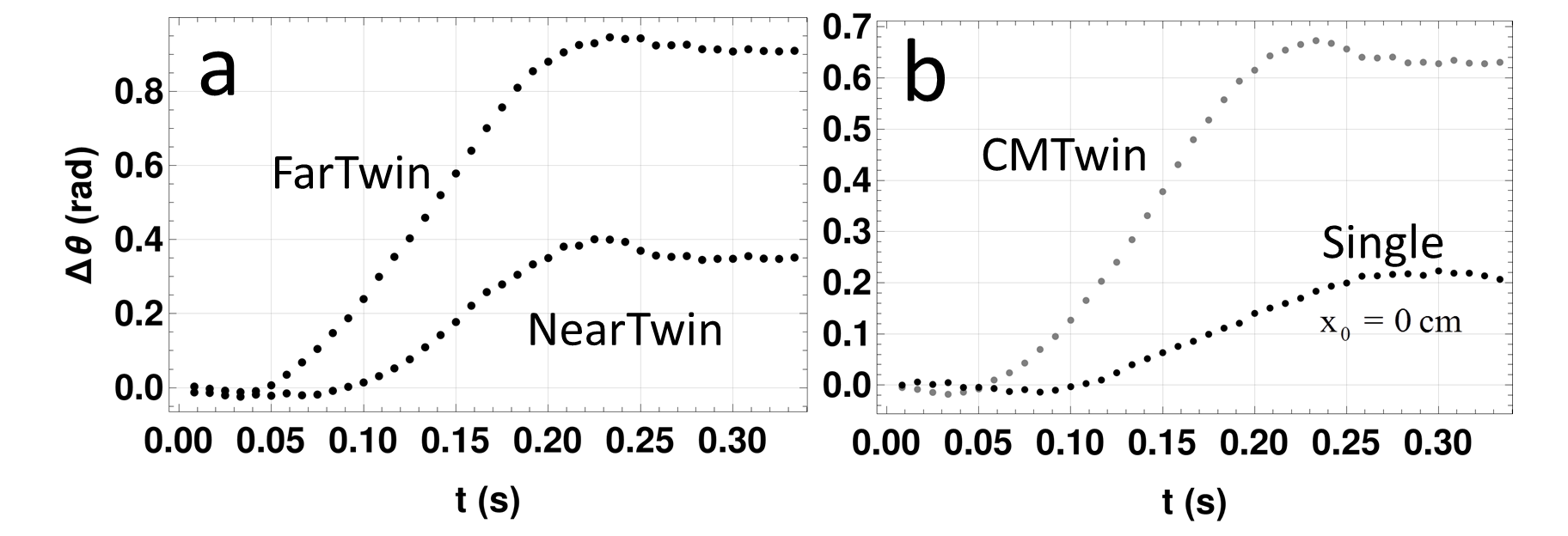}
%\vspace{3px}
%\includegraphics[width=230px]{Ac.eps}
%\centering
\caption{Rotation. (a) Time evolution of $\Delta \theta = \theta - \theta_0$ of the individual twin intruders after being released together. (b) In gray, time evolution of $\Delta \theta$ averaged between twin intruders. In black, rotation of a single intruder released from $x_0 = 0$ cm, i.e., the initial condition yielding maximum rotation. Trajectories averaged over 10 repetitions.}
\label{fig:Rotation}
\end{figure}

Finally, we end by briefly discussing the rotation of the twin intruders. Fig. \ref{fig:Rotation} (a) shows the rotation of each of the twin intruders, after averaging over 10 repetitions of the experiment. Both intruders rotate clockwise substantially more than the maximum rotation measured in single intruder experiments, which happens for $x_0=0$~cm \cite{Diaz-Melian2020}. Even more, the rotation averaged over both twins, shown in  Fig. \ref{fig:Rotation} (b), is also much larger than that for a single intruder released at $x_0=0$~cm. In \cite{Diaz-Melian2020}, we proposed a phenomenological model where the rotational acceleration, $\ddot{\theta}$, is proportional to the horizontal velocity of the center of mass, $\dot{x}$, and exponentially smaller as the penetration depth, $z$, increases. If we compare the average horizontal velocity and the vertical penetration of the twin intruders with the single intruder at $x_0 = 0\,cm$ as shown in Fig. \ref{fig:deltax} (b) and Fig. \ref{fig:deltaz} (b), respectively, the large difference observed in Fig. \ref{fig:Rotation} (b) can be explained in the light of the model.

% a substantially bigger angle that the  in the normal Notice that the NearTwin rotates approximately twice as much as a single intruder released at $x_0 = 0$ cm (see black dots in Fig. \ref{fig:Rotation} (b)), while it moves horizontally less than it. which is consistent with the fact that $\Delta x$ behaves quite similarly in both two cases. In fact, in \cite{Diaz-Melian2020} it was shown that $\Delta x$ and $\Delta \theta$ are intrinsically related (this fact is well illustrated when comparing figures \ref{fig:deltax} and \ref{fig:Rotation}: the onset of horizontal and angular motions start at the same time for each intruder). This idea also justifies the larger rotation of FarTwin seen in Fig. \ref{fig:Rotation} (a), which displayed a very large horizontal penetration (see Fig. \ref{fig:deltax}(a)). Finally, the gray dots in Fig. \ref{fig:Rotation} (b) indicate the rotation angle averaged between both twins shows, which is substantially larger than any rotation measured on a single intruder.

%The complex motion of the twin intruders is very difficult to understand in terms of the ``spring-bearded" intruder model. However, in the next section we will propose a concept that allows to establish a link between this case the ``cooperative dynamics" observed in sets of intruders released far from any boundaries.

\section*{Conclusion}

%We have fully characterized the motion of solid intruders penetrating a granular bed near a wall, after being released on the free granular surface at a distance $x_0$ from the vertical boundary. Individual cylindrical intruders are repelled by the wall in a length that decreases with $x_0$: if $x_0$ is larger than the characteristic length of the force chains $l$ , the intruder does not ``feel" the effect of the wall (the relation of $l$ with the dimensions of the experimental setup will be the subject of future work). This distancing from the wall was reproduced by modeling the force chains associated to the penetrating object as springs of finite length attached to the intruder, which interact repulsively with the vertical boundary.

%Our experiments also reveal that the interaction with the wall produces rotation of the intruder around its symmetry axis, which is clockwise or counter-clockwise for small or large values of $x_0$, respectively.

We have observed non-trivial cooperative motion when two intruders are released side-by-side from a vertical wall: the intruder nearer to the boundary ``transmits" the repulsion to the second one, resulting in an unexpectedly large lateral motion. Discrete Elements Simulations show that when four intruders arranged horizontally are released from the free surface near the wall, they are repelled from it by mechanisms qualitatively analogous to those operating on the twin intruders. The observed phenomena tend to reinforce the analogy between the behaviour of a set of intruders released far from the walls, and half of them released near one wall: the repulsion by the boundary seems equivalent to that expected from ``image" intruders on the other side of the wall.

Discrete element simulations also show that a chain of four intruders released vertically near a wall are ``bent away" from it, resulting in a very large repulsion of the intruder originally near the free granular surface.

Finally, twin intruders not only experience very large repulsion, but very large rotation, as compared to isolated intruders. This phenomenon can be explained by the relatively large horizontal velocity of the twin intruders, which provokes the frictional gradient responsible for the rotation.

These findings may shed light on engineering and geophysical scenarios involving ---for example-- the sedimentation of rocks near vertical walls.

%Finally, we have established a link between all the observations described before and previous experiments on intruders released far from any walls by introducing the concept of the {\it image granular intruder}. It can be defined as follows: the scenario of a set of intruders released near a wall is equivalent to one in which the wall is removed, and substituted by ``image" intruders. We believe that this concept may become a useful tool to understand the role of boundaries in granular matter.

\section*{Acknowledgment}

We acknowledge the University of Havana's institutional project ``Granular media: creating tools for the prevention of catastrophes". The Institute ``Pedro Kourí" is thanked for allowing us using their computing cluster. E. Altshuler found inspiration in the late M. \'Alvarez-Ponte.

\section*{Compliance with ethical standards}
Conflict of Interest: The authors declare that they have no conflict of interest.

% Bibliography
\bibliographystyle{unsrt}
\bibliography{biblio}

\end{document}

\title{Supplemental Material for \emph{Intruders cooperatively interact with a wall into granular matter}}

\author[a]{M. Espinosa$^*$}
\author[a,b]{V. L. D\'{\i}az$^*$}
\author[a]{A. Serrano-Muñoz}
\author[a]{E. Altshuler}

\affil[a]{Group of Complex Systems and Statistical Physics, Physics Faculty, University of Havana, 10400 Havana, Cuba}
\affil[b]{The Soft and Complex Materials Lab, IST Austria, 3400 Klosterneuburg, Austria}
\affil[*]{Equally contributed to the article}

% Please give the surname of the lead author for the running footer
        % <-this % stops a space
% The paper headers

\maketitle

\thispagestyle{empty}

\section*{Materials and Methods}

Expanded polyestirene particles of density $0.014\pm$ $0.002$\,g/cc and diameter distributed between $2.0$ and $6.5$\,mm, peaking at $5.8$\,mm were deposited into a Hele-Shaw cell of width $47.5$\,cm, height $60$\,cm and thickness $5.3$\,cm. The cell consisted in a three-piece aluminum frame and two glass plates, as shown in Fig. 4(a) of the main text. Cylindrical objects of height $5.2$\,cm and diameter $7.5$\,cm were used as intruders. The penetration process was followed using a GoPro camera with a resolution of $720\times 1280$\,pixels, at $120$ frames per second, that took videos through one of the glass faces. 

Two colored dots situated at the center and near the border of one circular face of the cylinders served as reference points for image analysis. Using them, the motion of the intruder's center of mass could be tracked within an uncertainty of $0.16$\,mm, and the angle of rotation of the intruder around its symmetry axis could be measured within an uncertainty of $0.005$\,rad. Between experiments, the granular material was removed from the cell, and then it was refilled following a precise protocol: using a specially designed funnel with a rectangular cross-section of $2.5\times19.5$\,cm, the granular material was gently deposited from the bottom to a height of approximately $40$\,cm inside the cell, as the funnel was slowly elevated. This resulted in a packing fraction of $0.65 \pm 0.01$.

\begin{figure}[!b]
\includegraphics[width=230px]{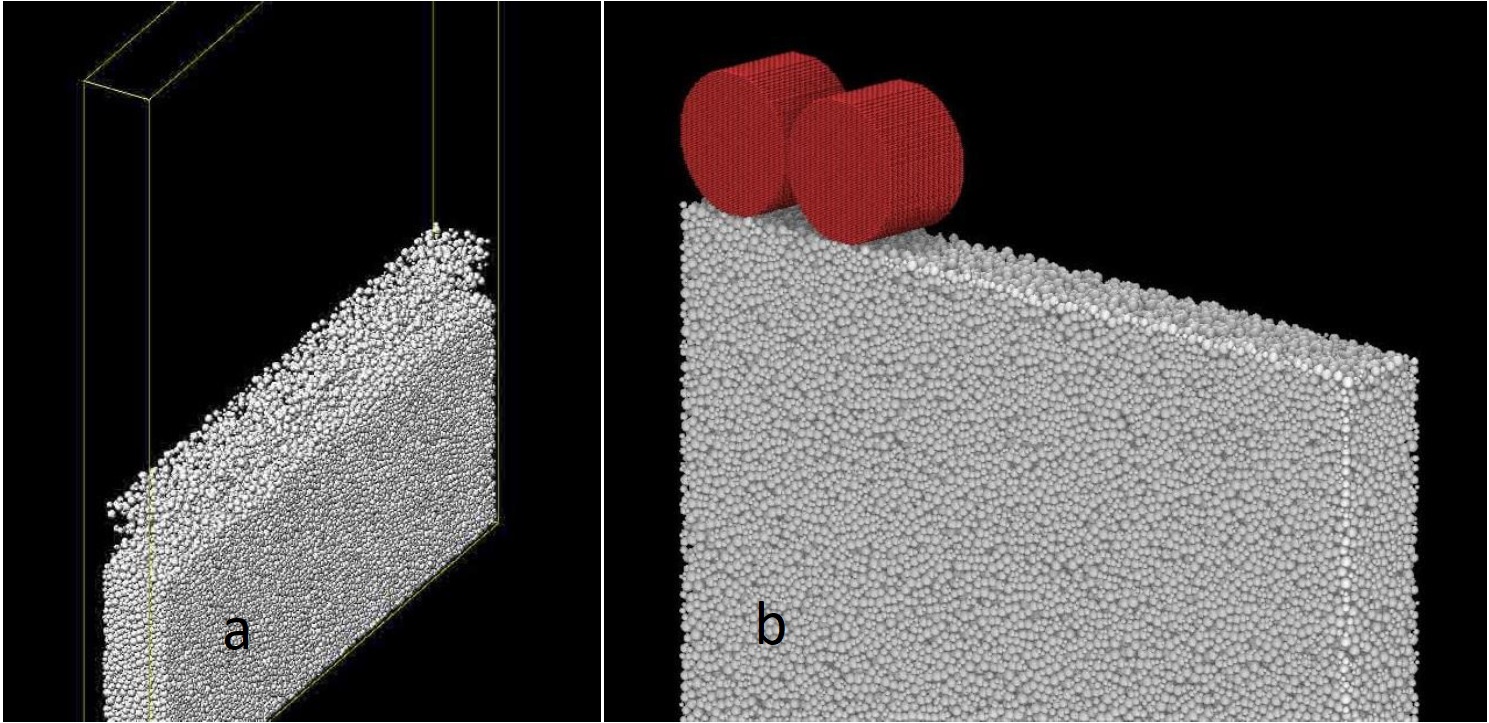}
%\vspace{3px}
%\includegraphics[width=230px]{Ac.eps}
%\centering
\caption{Simulation rendered images. (a) Preparation of the granular bed systematically and uniformly pouring batches of particles. The upper layer of less densely packed grains are particles that are still almost free-falling on top of the granular bed. (b) Initial step of the release of the twin intruders into the granular bed.}
\label{fig:SimSet}
\end{figure}

\begin{figure}[!b]
\includegraphics[width=230px]{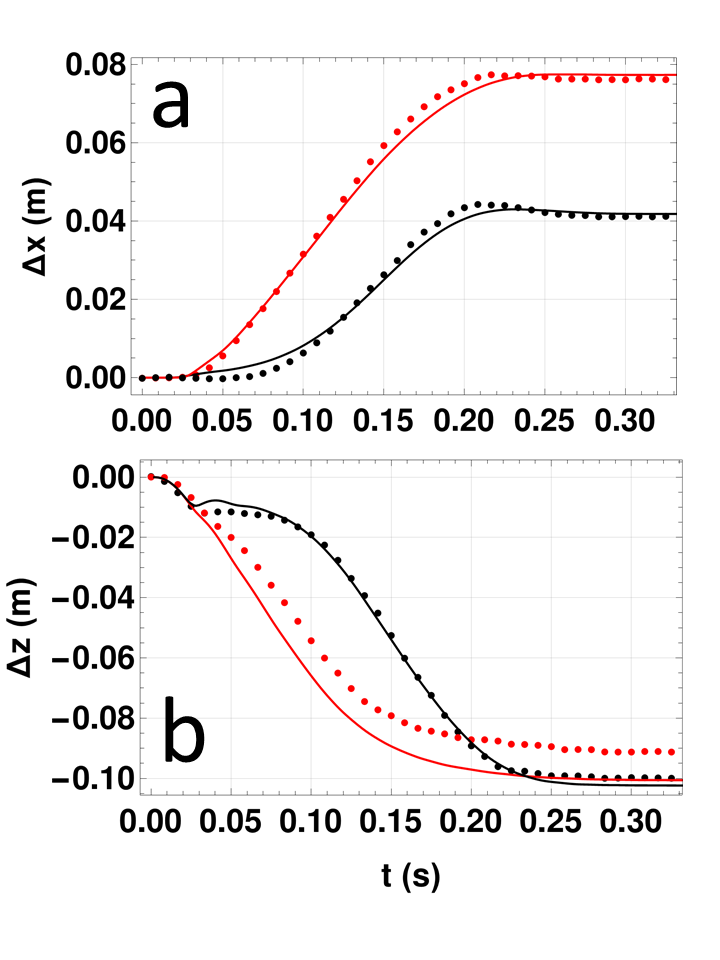}
%\vspace{3px}
%\includegraphics[width=230px]{Ac.eps}
%\centering
\caption{DEM simulation results vs experimental values. (a) The continuous line represents the horizontal displacement of the twin intruders (NearTwin:black, FarTwin:red) as an average over 10 numerical simulations of the experiment. The dotted line corresponds to Fig. 4 (a) of the main text. (b) The continuous line represents the vertical penetration of the twin intruders. The dotted line corresponds to Fig. 5 (a) of the main text.}
\label{fig:SimXZ}
\end{figure}

\section*{DEM simulations}

The DEM simulations were done using LAMMPS \cite{plimpton2007lammps} as described in the main text. The interaction between particles is ruled by a Hertzian contact model \cite{brilliantov1996model,silbert2001granular,zhang2005jamming} using the following parameters:

\begin{itemize}
	\item Elastic constant for normal contact: $K_n = 3.6\times10^5$.
	\item Elastic constant for tangential contact: $K_t = 5.5\times10^5$.
	\item Viscoelastic damping constant for normal contact: $\gamma_n = 6.8\times10^6$.
	\item Viscoelastic damping constant for tangential contact: $\gamma_t = \gamma_n/2$.
\end{itemize}

The constants were calculated using the material properties \cite{yu2005improved,iliuta2003concept}:

\begin{itemize}
	\item Young Modulus: $E = 5\,GPa$.
	\item Restitution coefficient: $e = 0.1$.
	\item Poisson's ratio: $\nu = 2/7$.
	\item Friction coefficient: $\mu = 0.5$.
	\item Particle density: $\rho = 0.014$\,g/cc.
\end{itemize}

The intruders are modeled as many-particle rigid bodies with density $1063$\,g/cc. A time-step of $5.6\times10^{-7}\,s$ was selected for the simulation based on the collision duration \cite{luding1998collisions}. Fig. \ref{fig:SimSet} shows rendered images of the system configuration during the preparation and release of the twin intruders.

The DEM simulations were able to reproduce the experimental results over an average of 10 simulations with different random seeds for the particle generation, within root-mean-square errors (RMSE) of 0.0417, 0.0276, 0.0226 and 0.117 (normalized to the maximum value of each experimental curve) for the NearTwin and FarTwin in Fig. \ref{fig:SimXZ} (a) and (b) respectively. Fig. \ref{fig:SimXZ} (a) shows a good fit for the lateral repulsion of the intruders, while Fig. \ref{fig:SimXZ} (b) shows a slight discrepancy in the vertical penetration of the FarTwin, as expressed by the relatively high RMSE. It may be caused by “quenched” effects, such as the inhomogeneous deposition of grains in the cell associated with our filling protocol. In the simulations the uniform pour does not account for this effect.

The simulations were not able to reproduce quantitatively the rotation of the twin intruders; only qualitatively (FarTwin rotates much more than the NearTwin). This can be related to certain coefficients of which we have no knowledge and so were not reproduced in the DEM simulations, such as the friction between the intruders and the particles (presumably an important factor in the rotation dynamic as explained in \cite{Diaz-Melian2020}). 

\markboth {Mespinosa \MakeLowercase{\textit{et al.}}: Intruders cooperatively interact with a wall into granular matter}{Mespinosa \MakeLowercase{\textit{et al.}}: Intruders cooperatively interact with a wall into granular matter}

% Bibliography
\bibliographystyle{unsrt}
\bibliography{biblio}

\vfill